\documentclass[runningheads]{llncs}
\usepackage[T1]{fontenc}
\usepackage{amsmath}
\usepackage{amssymb}
\usepackage{booktabs}
\usepackage{multirow}
\usepackage[breaklinks,colorlinks]{hyperref}
\usepackage{breakcites}

\usepackage{array}
\newcolumntype{P}[1]{>{\centering\arraybackslash}p{#1}}

\usepackage[capitalize]{cleveref}
\crefname{section}{Sec.}{Secs.}
\Crefname{section}{Section}{Sections}
\Crefname{table}{Table}{Tables}
\crefname{table}{Tab.}{Tabs.}
\usepackage{graphicx}
\def\equationautorefname{Eq.}
\def\sectionautorefname{Sec.}
\def\tableautorefname{Tab.}
\def\figureautorefname{Fig.}
\def\appendixautorefname{Appx.}

\begin{document}
\title{HealthyGAN: Learning from Unannotated Medical Images to Detect Anomalies Associated with Human Disease}
\titlerunning{HealthyGAN}
%
\author{Md Mahfuzur Rahman Siddiquee\inst{1,2} \and
Jay Shah\inst{1,2} \and
Teresa Wu\inst{1,2} \and
Catherine Chong\inst{2,3} \and
Todd Schwedt\inst{2,3} \and
Baoxin Li\inst{1,2}}
\authorrunning{M. Rahman Siddiquee et al.}
%
\institute{Arizona State University, Tempe, AZ, USA \\
\email{mrahmans@asu.edu}
\and
ASU-Mayo Center for Innovative Imaging, Tempe, AZ, USA \and
Mayo Clinic, Phoenix, AZ, USA}
\maketitle              

\vspace{-240pt}
\begin{center}
    \centering
    \textcolor{blue}{Please cite the paper as M.M. Rahman Siddiquee, J. Shah, T. Wu, C. Chong, T. Schwedt, and B. Li. HealthyGAN: Learning from Unannotated Medical Images to Detect Anomalies Associated with Human Disease. International Workshop on Simulation and Synthesis in Medical Imaging, 2022.}
\end{center}

\vspace{170pt}

\begin{abstract}
Automated anomaly detection from medical images, such as MRIs and X-rays, can significantly reduce human effort in disease diagnosis. Owing to the complexity of modeling anomalies and the high cost of manual annotation by domain experts (e.g., radiologists), a typical technique in the current medical imaging literature has focused on deriving diagnostic models from healthy subjects only, assuming the model will detect the images from patients as outliers. However, in many real-world scenarios, unannotated datasets with a mix of both healthy and diseased individuals are abundant. Therefore, this paper poses the research question of how to improve unsupervised anomaly detection by utilizing (1) an unannotated set of mixed images, in addition to (2) the set of healthy images as being used in the literature. To answer the question, we propose HealthyGAN, a novel one-directional image-to-image translation method, which learns to translate the images from the mixed dataset to only healthy images. Being one-directional, HealthyGAN relaxes the requirement of cycle-consistency of existing unpaired image-to-image translation methods, which is unattainable with mixed unannotated data. Once the translation is learned, we generate a difference map for any given image by subtracting its translated output. Regions of significant responses in the difference map correspond to potential anomalies (if any). Our HealthyGAN outperforms the conventional state-of-the-art methods by significant margins on two publicly available datasets: COVID-19 and NIH ChestX-ray14, and one institutional dataset collected from Mayo Clinic. The implementation is publicly available at~\url{https://github.com/mahfuzmohammad/HealthyGAN}.

\keywords{Anomaly detection \and COVID-19 detection \and Thoracic disease detection \and Migraine detection \and Image-to-Image Translation.}
\end{abstract}

\begin{figure}
    \centering
    \includegraphics[width=\linewidth]{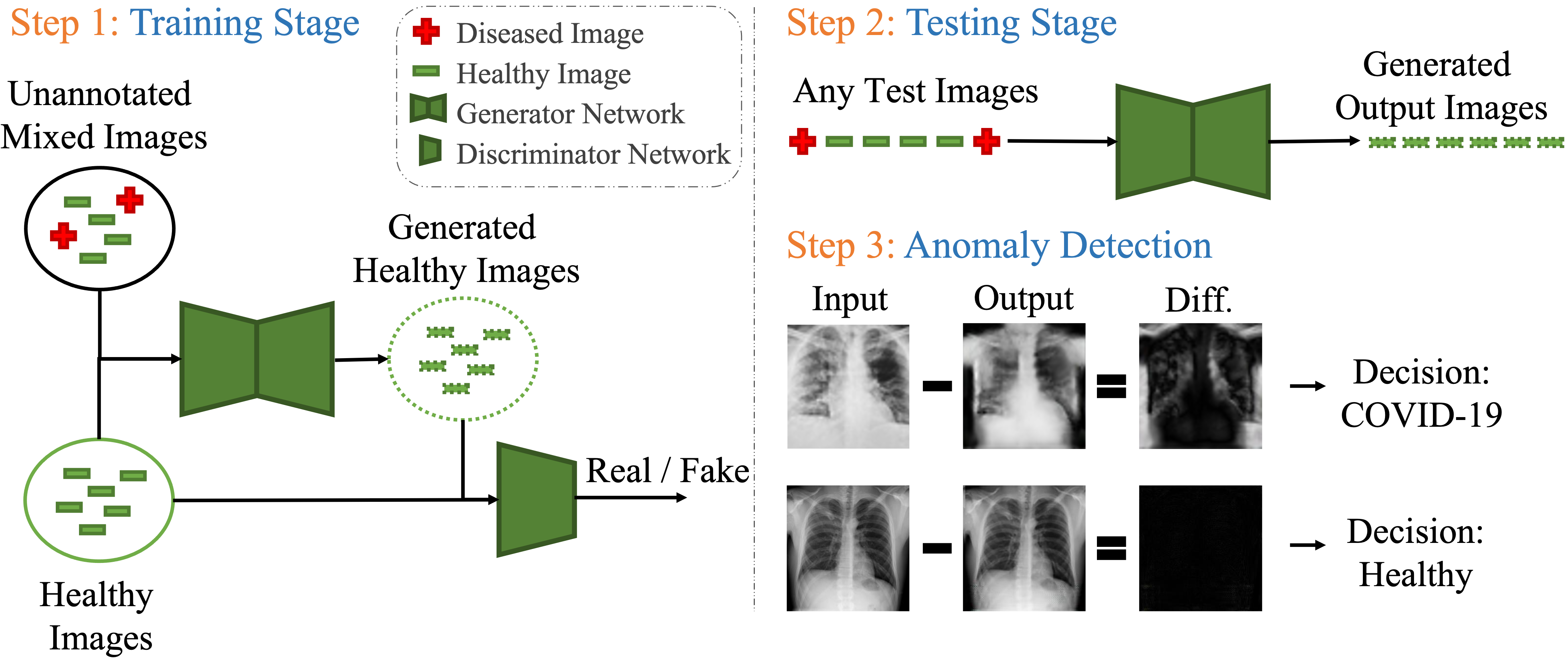}
    \caption{Overview of the proposed anomaly detection method. At the training stage, our proposed HealthyGAN learns to generate healthy images utilizing an unannotated dataset mixed with both healthy and potential diseased/anomalous images, in addition to, a set of healthy images. At the testing stage, the absolute difference between the translated healthy image and the input image reveals the presence of an anomaly.}
    \label{fig:anomaly_detection_overview}
\end{figure}

\section{Introduction}
\label{sec:introduction}
Supervised learning from a large annotated dataset is becoming easier~\cite{he2015delving,esteva2017dermatologist}, due to deep neural networks. For problems like anomaly detection (e.g., rare disease detection in medical images), however, it may often be challenging to obtain large enough datasets of annotated samples, making it impractical to rely on supervised learning for the task. Therefore, many recent attempts to develop diagnostic models learn only from the images of healthy (i.e. normal) subjects ~\cite{chen2018unsupervised,schlegl2017unsupervised,schlegl2019f,alex2017generative,zenati2018efficient,zenati2018adversarially,akcay2018ganomaly,gherbi2019encoding,roth2021towards,defard2021padim,sabokrou2018adversarially}. However, in practice, {\em unannotated} anomalous samples (mixed with normal samples) are usually available and what is missing is the elaborated annotation. In this research, we seek to answer the question: {\em How can we utilize an unannotated mixed dataset, in addition to the set of normal images, to improve the performance of anomaly detection?}


The answer to the question has been explored in the distant past for lesion detection in vascular CT images using SVM~\cite{zuluaga2011learning}. To the best of our knowledge, we have not seen any deep learning approach designed to address this question. Therefore, aiming to achieve a more generalized solution, in this paper we have developed a novel one-directional unpaired image-to-image translation network, termed HealthyGAN, based on Generative Adversarial Network~\cite{goodfellow2014generative,goodfellow2020generative} (GAN). The proposed HealthyGAN learns to translate any images from a mixed dataset to normal images (a.k.a. healthy images) and detect abnormalities based on the differences between the input and the output images as illustrated in~\figureautorefname~\ref{fig:anomaly_detection_overview}. We want to highlight that existing unpaired image-to-image translation methods~\cite{liu2017unsupervised,shen2017learning,yi2017dualgan,zhu2017unpaired,zhu2017toward,choi2017stargan,mejjati2018unsupervised,zhang2018generative,he2019attgan,liu2019stgan,zhao2020unpaired} are not suitable for solving this problem since they are (1) bi-directional, requiring both abnormal-to-normal and normal-to-abnormal translation to ensure cycle-consistency~\cite{zhu2017unpaired}; or (2) require the images to be annotated as normal vs. abnormal {\em prior} to training. However, the translation of normal images back to abnormal images is not a feasible approach while using unannotated datasets. To address this challenge, HealthyGAN employs two important properties for improving anomaly detection: (1) unpaired image-to-image translation; and (2) one-directional image-to-image translation. To achieve these properties, we introduce a novel reconstruction loss that ensures effective cycle-consistency during the one-directional translation. Specifically, unlike traditional cycle-consistency loss~\cite{zhu2017unpaired}, our reconstruction loss utilizes learned attention-masks to generate the reconstructed images for cycle-consistency. Since all the image manipulation for backward-cycle occurs using basic mathematical operations, there is no need for image annotation (see~\sectionautorefname~\ref{sec:method}).

Through extensive experiments, we demonstrate that HealthyGAN outperforms existing state-of-the-art anomaly detection methods by significant margins on two public datasets: COVID-19 and NIH ChestX-ray14; and one single institute dataset for Migraine detection. This performance is attributed to HealthyGAN's capability of utilizing unannotated diseased/anomalous images during training. In summary, we make the following contributions:
\begin{itemize}
    \item We introduce a novel one-directional unpaired image-to-image translation method for anomaly detection that utilizes unannotated mixed datasets with images from healthy subjects and patients.
    \item We develop a novel reconstruction loss for ensuring cycle-consistency without requiring annotated inputs. 
    \item With three challenging medical datasets, we perform extensive experiments comparing the proposed method, HealthyGAN, against the conventional state-of-the-art anomaly detection methods, and we report significant performance improvements and provide detailed analysis.
\end{itemize}

\section{HealthyGAN: The Proposed Method}
\label{sec:method}

\subsection{Network Architecture}

The proposed HealthyGAN consists of a discriminator network and a generator network. The discriminator network follows PatchGAN~\cite{isola2017image,li2016precomputed,zhu2017unpaired} architecture and is similar to the ones used in~\cite{choi2017stargan,siddiquee2019learning}. Our discriminator distinguishes whether the input image is a real or a fake (i.e. generated) healthy image.

The generator network takes any images without knowing their labels and translates them to only healthy images. For training, we use a mixed dataset (Set $A$) containing both diseased and healthy images, and another dataset (Set $B$) containing only healthy images. The generated images of these corresponding Sets are denoted as $A'$ and $B'$, respectively. The generator does not generate the $A'$ and $B'$ images directly; rather, it generates intermediate healthy images $B_{int}$ and masks $M$. The masks' values are in the range [0 -- 1], where 0 means background pixel, and 1 means foreground pixel. Then we produce the final generated image $B'$ following~\equationautorefname~\ref{eq:mask_generate_b} and similarly, $A'$ following~\equationautorefname~\ref{eq:mask_generate_a}.

\begin{equation}
    \label{eq:mask_generate_b}
    B' = B_{int} \odot M + A \odot (\mathbf{1} -  M)
\end{equation}
\begin{equation}
    \label{eq:mask_generate_a}
    A' = A \odot M + B_{int} \odot (\mathbf{1} -  M)
\end{equation}

\begin{figure*}[t!]
    \centering
    \includegraphics[width=\linewidth]{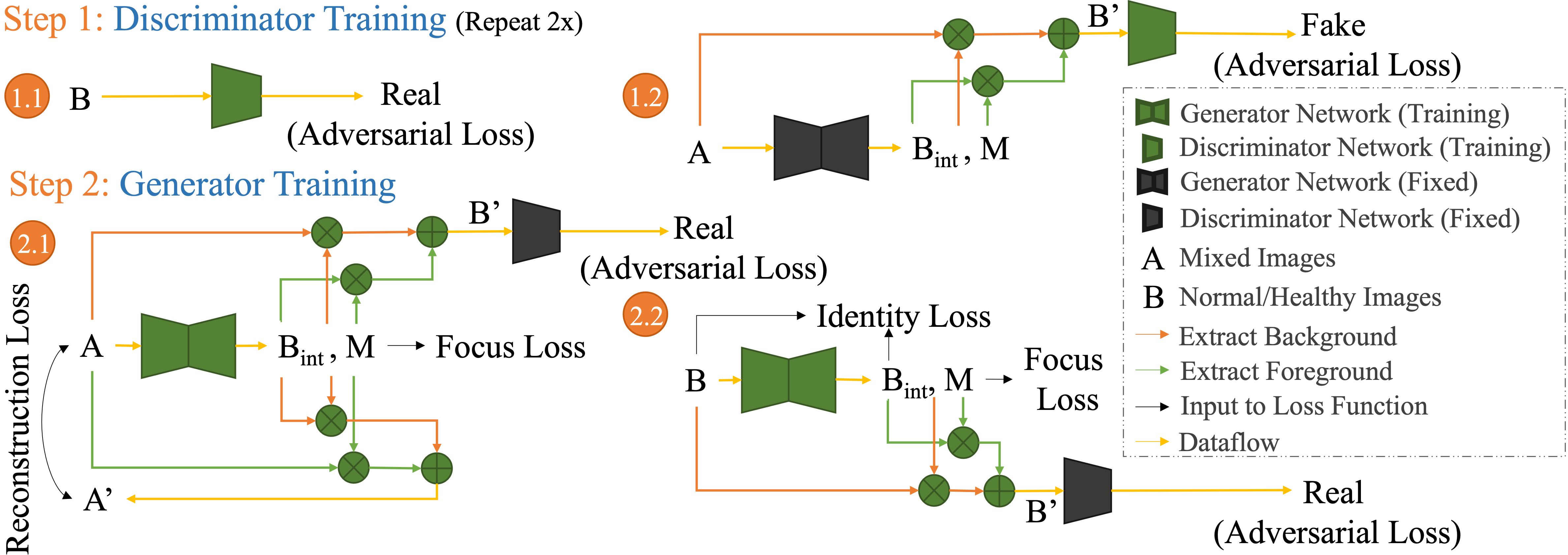}
    \caption{A single training iteration of HealthyGAN, each of which consists of 2 sub-steps: (1) discriminator training and (2) generator training. The discriminator learns to distinguish real healthy images from the generated ones while the generator tries to fool the discriminator by generating realistic healthy images. Please see~\sectionautorefname~\ref{sec:method} for training details,~\appendixautorefname~\ref{appsec:implementation_details} for hyper-parameters, and~\appendixautorefname~\ref{appsec:arch} for the network architectures.}
    \label{fig:healthy_gan_training}
\end{figure*}

\noindent If the image from Set $A$ is diseased, we expect the mask $M$ to activate the diseased region as foreground; otherwise, we expect $M$ to be empty/zero. It is worth noting the similarity between~\equationautorefname~\ref{eq:mask_generate_a} and the cycle-consistency concept introduced in~\cite{zhu2017unpaired}. Since the proposed method is controlling the image generation, partially, by the mask $M$, it neither requires a label nor an additional generator network to generate $A'$ (\figureautorefname~\ref{fig:healthy_gan_training}). As the generator network translates the input images to a single direction, we call it a one-directional image-to-image translation method.

\subsection{Training}
\label{subsec:method_training}

\figureautorefname~\ref{fig:healthy_gan_training} depicts the detailed training methodology of HealthyGAN. We train the generator and the discriminator network, alternately, like any GAN model. At each training step, we update the weights of the generator once for every two weight updates of the discriminator network and repeat until convergence.

The discriminator is trained to learn the healthy images in $B$ as real and any images from the generator to be fake using an {\em adversarial loss} defined in~\equationautorefname~\ref{eq:adv_d_loss}.  

\vspace{-0.3cm}
\begin{equation}
\label{eq:adv_d_loss}
\begin{split}
\mathcal{L}_{adv}^{D} = \thinspace & ~\mathbb{E}_{x \in A}[D_{real/fake}(G(x))] - \mathbb{E}_{x \in B}[D_{real/fake}(x)]\\
&+ {\lambda}_{gp} \thinspace {\mathbb{E}}_{\hat{x}}[{{(||{\triangledown}_{\hat{x}} {D}_{real/fake}(\hat{x})||}_{2} - 1)}^{2}]
\end{split}
\end{equation}

\noindent Here, $G(x)$ denotes the output of the generator and is obtained by~\equationautorefname~\ref{eq:mask_generate_b}. $D_{real/fake}(x)$ denotes the output of the discriminator network. \equationautorefname~\ref{eq:adv_d_loss} is the revised adversarial loss based on the Wasserstein GAN~\cite{arjovsky2017wasserstein} and an added gradient penalty~\cite{gulrajani2017improved} with weight ${\lambda}_{gp}$ which helps to stabilize the training.

The objective of the generator is to translate any input image to the corresponding healthy image. To be specific, if the input image is a healthy image, the generator is expected to behave like an autoencoder. If the input is a diseased image, the generator should remove anomalous parts and produce a healthy image in the output. The {\em adversarial loss} for the generator is defined in~\equationautorefname~\ref{eq:adv_g_loss}.

\vspace{-0.2cm}
\begin{equation}
\label{eq:adv_g_loss}
    \mathcal{L}_{adv}^{G} = -\sum_{x \in \{A, B\}} \mathbb{E}_{x}[D_{real/fake}(G(x))]
\end{equation}

\noindent For the known healthy image set, $B$, the generator should behave like an autoencoder. Hence, we apply an {\em identity loss} (defined in~\equationautorefname~\ref{eq:id_loss}) for these images.

\vspace{-0.2cm}
\begin{equation}
\label{eq:id_loss}
    \mathcal{L}_{id} = \mathbb{E}_{x \in B}[||G_{int}(x) - x||_1]
\end{equation}

\noindent Here, $G_{int}$ denotes the generated images before applying the masks ($B_{int}$ in~\figureautorefname~\ref{fig:healthy_gan_training}). Since we train HealthyGAN using unpaired images we add a {\em reconstruction loss} (\equationautorefname~\ref{eq:rec_loss}) to ensure that the generated images are close to the input images.

\vspace{-0.5cm}
\begin{equation}
\label{eq:rec_loss}
\begin{split}
    \mathcal{L}_{rec} =\ & \mathbb{E}_{x \in A, y \in A'}[||x - y||_1]
\end{split}
\end{equation}

\noindent To control the size of the masks, we have adopted the {\em focus loss} of~\equationautorefname~\ref{eq:focus_loss} from~\cite{nizan2020breaking}.

\vspace{-0.5cm}
\begin{equation}
\label{eq:focus_loss}
\begin{split}
    \mathcal{L}_{f} =\ & {\lambda}_{fs} (\sum_{i = 1}^{n} M_i / n)^2 + {\lambda}_{fz} \frac{1}{n} \sum_{i = 1}^{n} \frac{1}{|M_i - 0.5| + \epsilon}
\end{split}
\end{equation}

\noindent Here, $n$ denotes the number of pixels in the mask $M$ and $M_i$ denotes a pixel in it. The first component controls the size of the mask and the second component forces the values to be close to 0/1. ${\lambda}_{fs}$ and ${\lambda}_{fz}$ are relative weights of these components, respectively.

Combining all losses, the final full objective function for the discriminator and generator can be described by~\equationautorefname~\ref{eq:d_full_loss} and~\equationautorefname~\ref{eq:g_full_loss}, respectively.

\vspace{-0.2cm}
\begin{equation}
\label{eq:d_full_loss}
    \mathcal{L}_D = \mathcal{L}_{adv}^{D}
\end{equation}
\vspace{-0.2cm}
\begin{equation}
\label{eq:g_full_loss}
    \mathcal{L}_G = \mathcal{L}_{adv}^{G} + \lambda_{rec}\mathcal{L}_{rec} +  \lambda_{id}\mathcal{L}_{id} + \lambda_{f}\mathcal{L}_{f}
\end{equation}
where $\lambda_{rec}$, $\lambda_{id}$, and $\lambda_{f}$ determine the relative importance of the \textit{reconstruction loss}, \textit{identity loss}, and \textit{focus loss}, respectively.

\subsection{Detecting Anomalies}

\figureautorefname~\ref{fig:anomaly_detection_overview} provides an overview of the proposed anomaly detection method. Given an unannotated mixed dataset containing a mixture of both diseased and healthy images $A$ and another dataset containing only healthy images $B$, we train the HealthyGAN as described in~\sectionautorefname~\ref{subsec:method_training}. Once trained, we first translate each of the test images into healthy images, and then we compute the absolute difference between the generated healthy images and the input images. We expect the resultant difference images to show the diseased regions if the input is a diseased image; otherwise, we expect the difference image to contain only pixels with a value of zero or very close to zero. Therefore, we detect the presence of the disease by checking the mean value of the difference images. Please note that the difference images indicate the presence of the disease/anomaly, and can also serve to localize image regions that associate with the disease/anomaly. However, the proposed HealthyGAN does not guarantee the detection of all the disease-specific features. Identifying only a subset of the disease-specific features is sufficient to serve the purpose of this study.

\section{Experiments and Results}
\label{sec:experiments}

\noindent\textbf{Competing Methods.}~We have compared the proposed HealthyGAN with 6 state-of-the-art anomaly detection methods currently in use. We have selected these methods as they are the most recent. Among them, ALAD~\cite{zenati2018adversarially}, ALOCC~\cite{sabokrou2018adversarially}, f-AnoGAN~\cite{schlegl2019f}, and Ganomaly~\cite{akcay2018ganomaly} are methodologically the closest to the proposed HealthyGAN. We have excluded other methodologically similar works such as EGBAD~\cite{zenati2018efficient} and AnoGAN~\cite{schlegl2017unsupervised} from our competing methods' list since ALAD and f-AnoGAN are improved versions of these methods, respectively. However, we have included PatchCore~\cite{roth2021towards} and PaDiM~\cite{defard2021padim}, though methodologically different than the proposed HealthyGAN, as they are state of the art for novelty detection in natural image dataset like MVTec AD~\cite{bergmann2019mvtec,bergmann2021mvtec}. 

\noindent\textbf{Evaluation.} We have compared the proposed HealthyGAN for anomaly detection with the conventional methods using the AUC score from the receiver operating characteristic (ROC) curve. In addition, we have reported precision, recall, specificity, and F1 scores. To get the prediction score for HealthyGAN, we first take the absolute difference between the input image and its translated image. Then we compute the mean value of the resultant difference image. We found the mean to be more robust than the maximum. For the conventional methods, we have used the anomaly score generation method proposed by their corresponding authors.

\subsection{COVID-19 Detection}
\label{subsec:experiment_covid}

\noindent\textbf{Dataset.} We have utilized the COVIDx dataset from~\cite{Wang2020}. The original dataset contains a training set with 15,464 Chest X-rays (1,670 COVID-19 positives, 13,794 healthy) and a testing set with 200 Chest X-rays (100 positives and 100 healthy). For our experiments, we have randomly taken 10,031 healthy images for the known healthy training set. For the mixed unannotated training set, we have randomly taken 3,663 healthy and 1,570 COVID-19 positive images.

\noindent\textbf{Results.} The top section in~\tableautorefname~\ref{tab:anomaly_detection_results} summarizes the COVID-19 detection results. As seen, HealthyGAN achieves COVID-19 detection AUC of 0.84 outperforming all the conventional methods by a large margin. It also achieves the best precision, specificity, and F1 scores of 0.76. In contrast, f-AnoGAN, the top performing among the competing methods, achieves an AUC score of only 0.64 which is 0.20 points lower than the proposed HealthyGAN. f-AnoGAN achieves precision, recall, specificity, and F1 scores of 0.55, 0.53, 0.56, and 0.54, respectively.~\figureautorefname~\ref{fig:results} shows qualitative results of COVID-19 detection by HealthyGAN.

\subsection{Chest X-ray 14 Diseases Detection}
\label{subsec:experiment_xray14}

\noindent\textbf{Dataset.} We have utilized only the Posterior Anterior (PA) X-rays from the ChestX-ray14 dataset~\cite{wang2017chestx} for this experiment. The dataset contains X-rays with one or more of 14 thoracic diseases. For ease of evaluation, we selected the X-rays having only one disease. From the resultant X-rays, we used 10,000 healthy X-rays for the known healthy training set; 5,000 healthy X-rays, and 3,195 diseased X-rays for the mixed unannotated training set. For the validation set, we used 4,000 healthy and 4,000  diseased X-rays and for the testing set, we used 10,000 healthy and 10,000 diseased X-rays.

\begin{figure*}[!t]
    \centering
    \includegraphics[width=\textwidth]{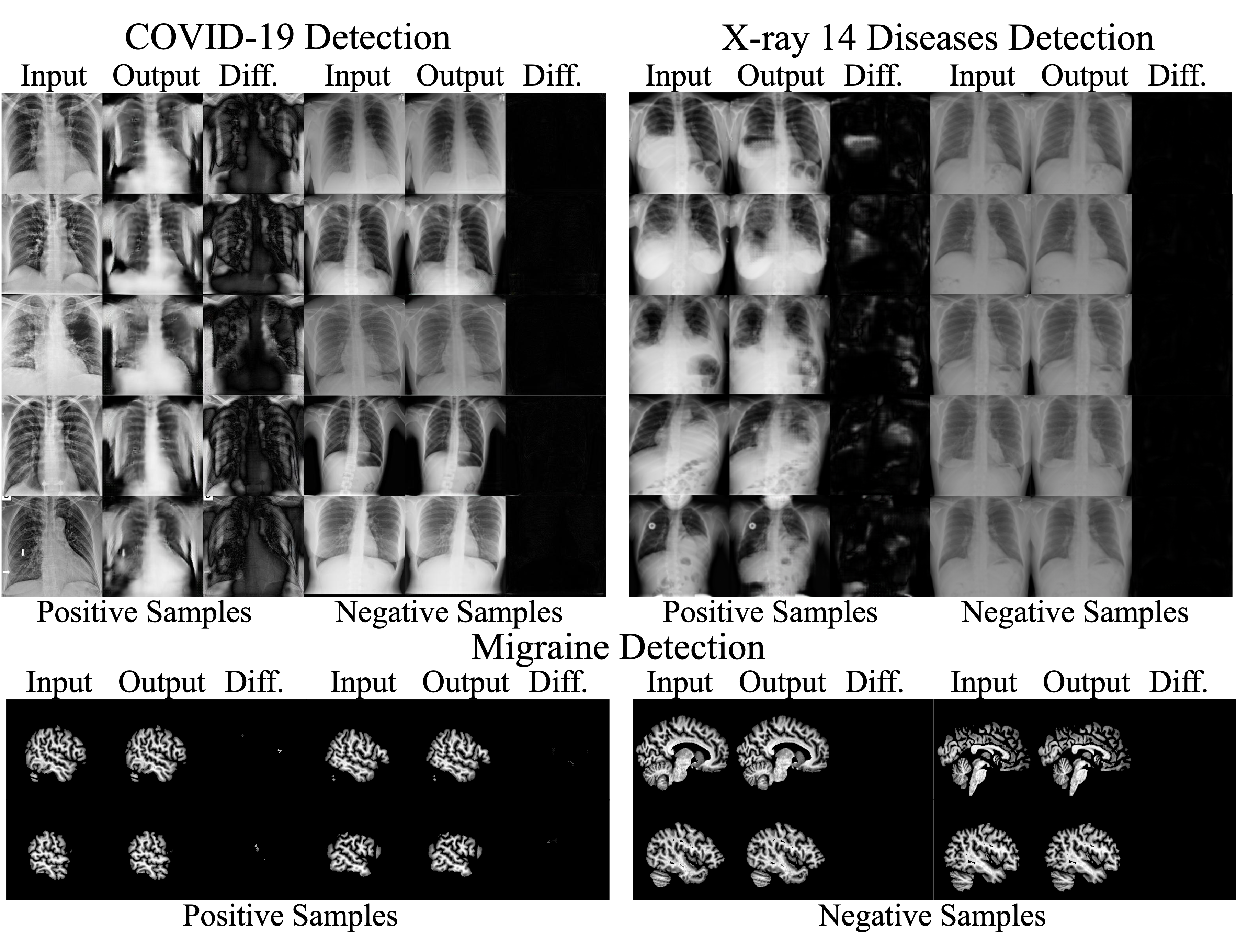}
    \vspace{-0.8cm}
    \caption{Qualitative results of COVID-19, Chest X-ray 14 diseases detection, and Migraine detection by HealthyGAN. As seen, HealthyGAN has resulted in a high response in the difference maps for positive samples compared to the negative samples.}
    \label{fig:results}
\end{figure*}

\noindent\textbf{Results.} The middle section of \tableautorefname~\ref{tab:anomaly_detection_results} summarizes the 14 diseases detection results. As seen, HealthyGAN achieves the best detection AUC score of 0.56 while f-AnoGAN performs the second best with an AUC score of 0.55. They both achieve precision, recall, specificity, and F1 scores of 0.55.~\figureautorefname~\ref{fig:results} contains qualitative results of the 14 diseases detection by HealthyGAN.

\begin{table}[!t]\centering
\caption{Summary of the anomaly detection results. We have compared HealthyGAN with 6 state-of-the-art anomaly detection methods using 5 metrics on 3 medical imaging datasets. The best results are in \textbf{bold} and the second best results are \underline{underlined}.}
\label{tab:anomaly_detection_results}
\begin{tabular}{|c|c|c|c|c|c|c|c|c|c|}

\hline
{\rotatebox[origin=c]{45}{\textbf{Datasets}}} & 
{\rotatebox[origin=c]{45}{\textbf{Metrics}}} & 
{\rotatebox[origin=c]{45}{\textbf{ALAD}}} & 
{\rotatebox[origin=c]{45}{\textbf{ALOCC}}} & 
{\rotatebox[origin=c]{45}{\textbf{f-AnoGAN}}} & 
{\rotatebox[origin=c]{45}{\textbf{Ganomaly}}} & 
{\rotatebox[origin=c]{45}{\textbf{Padim}}} & 
{\rotatebox[origin=c]{45}{\textbf{PatchCore}}} &
{\rotatebox[origin=c]{45}{\textbf{HealthyGAN}}} \\
\hline

\multirow{5}{1.5cm}{\centering\textbf{COVID-19}} & \textbf{AUC} &0.58 &0.63 &\underline{0.64} &0.58 &0.56 &0.52 &\textbf{0.84} \\
& \textbf{Prec.} &0.49 &\underline{0.63} &0.55 &0.59 &0.56 &0.52 &\textbf{0.76} \\
& \textbf{Rec.} &\textbf{0.89} &0.63 &0.53 &0.60 &0.56 &0.53 &\underline{0.76} \\
& \textbf{Spec.} &0.09 &\underline{0.63} &0.56 &0.59 &0.56 &0.51 &\textbf{0.76} \\
& F1 &\underline{0.64} &0.63 &0.54 &0.60 &0.56 &0.52 &\textbf{0.76} \\
\hline
\multirow{5}{1.5cm}{\centering\textbf{X-ray 14 diseases}} & \textbf{AUC} & 0.53 &0.48 &\underline{0.55} &0.49 &0.54 &0.53 &\textbf{0.56} \\
& \textbf{Prec.} &0.53 &0.48 &\textbf{0.55} &0.49 &\underline{0.54} &0.53 &\textbf{0.55} \\
& \textbf{Rec.} &0.53 &0.48 &\textbf{0.55} &0.49 &\underline{0.54} &0.53 &\textbf{0.55} \\
& \textbf{Spec.} &0.53 &0.48 &\textbf{0.55} &0.49 &\underline{0.54} &0.53 &\textbf{0.55} \\
& \textbf{F1} &0.53 &0.48 &\textbf{0.55}&0.49 &\underline{0.54} &0.53 &\textbf{0.55} \\
\hline
\multirow{5}{1.5cm}{\centering\textbf{Migraine}} & \textbf{AUC} &0.60 &0.40 &0.50 &\underline{0.70} &0.35 &0.60 &\textbf{0.75} \\
& \textbf{Prec.} &0.60 &0.40 &0.50 &\underline{0.70} &0.36 &0.60 &\textbf{0.78} \\
& \textbf{Rec.} &0.60 &0.40 &0.50 &\underline{0.70} &0.40 &0.60 &\textbf{0.70} \\
& \textbf{Spec.} &0.60 &0.40 &0.50 &\underline{0.70} &0.30 &0.60 &\textbf{0.80} \\
& \textbf{F1} &0.60 &0.40 &0.50 &\underline{0.70} &0.38 &0.60 &\textbf{0.74} \\
\hline

\end{tabular}
\end{table}

\subsection{Migraine Detection}
\label{subsec:experiment_migraine}

\noindent\textbf{Dataset.} Our migraine dataset, collected by our collaborators at Mayo Clinic, contains 96 brain MRIs of migraine patients and 104 brain MRIs of healthy participants. We randomly selected 10 migraine patients and 10 healthy participants for each of the validation and test sets. The rest of the 76 migraine patients and 84 healthy participants were used as the mixed unannotated training set. For the known healthy set, we used 424 participants from the IXI public dataset~\cite{ixi}.

\noindent\textbf{Results.} The last section of \tableautorefname~\ref{tab:anomaly_detection_results} summarizes the results for migraine detection. HealthyGAN outperforms all the conventional methods by achieving an AUC score of 0.75. It also achieves the best precision, recall, specificity, and F1 scores of 0.78, 0.70, 0.80, and 0.74, respectively. In contrast, the top performing conventional method, Ganomaly, achieves AUC, precision, recall, specificity, and F1 scores of only 0.70.~\figureautorefname~\ref{fig:results} shows qualitative results of HealthyGAN.




\section{Conclusion}

We have introduced a novel one-directional unpaired image-to-image translation method for anomaly detection from medical images, named HealthyGAN. We have devised a methodology to utilize an unannotated mixed dataset with both normal and anomalous images during the training of the proposed HealthyGAN. It has been possible due to the proposed novel reconstruction loss that ensures effective cycle-consistency without requiring input image annotations. Our extensive evaluation has demonstrated the proposed HealthyGAN's superiority over the existing state-of-the-art anomaly detection methods. The superior performance is attributed to HealthyGAN's capability of utilizing unannotated anomalous images during training.

\subsubsection{Acknowledgments.} This research has been supported by the United States Department of Defense W81XWH-15-1-0286 and W81XWH1910534, National Institutes of Health K23NS070891, National Institutes of Health - National Institute of Neurological Disorders and Stroke, Award Number 1R61NS113315–01, and Amgen Investigator Sponsored Study 20187183. We thank Arizona State University Research (ASURC) Computing for hosting our computing resources.


%
%
%
%

\newpage
\appendix

\section{Implementation Details}
\label{appsec:implementation_details}

\noindent We have resized the input images to $256\times256$ for the experiments on COVID-19 and Migraine detection in~\sectionautorefname~\ref{subsec:experiment_covid} and~\sectionautorefname~\ref{subsec:experiment_migraine}, respectively. For the X-ray 14 diseases detection in~\sectionautorefname~\ref{subsec:experiment_xray14}, we have resized the images to $128\times128$. We have set $\lambda_{gp}$ = 10, $\lambda_{id}$ = 1, $\lambda_{rec}$ = 1, $\lambda_{f}$ = 0.1, $\lambda_{fz}$ = 1, and $\lambda_{fs}$ = 1 for all the experiments. For COVID-19 and Migraine detection, we have used a batch-size of 16. For X-ray 14 diseases, we have used 32. We trained the models for 400,000 iterations. We have used Adam optimizer with a learning rate of $1e^{-4}$. The learning rate has been decayed for the last 100,000 iterations. Once trained, we have picked the best model using Fréchet inception distance (FID)~\cite{heusel2017gans,Seitzer2020FID}. The network architecture details are provided in~\appendixautorefname~\ref{appsec:arch}.



\section{Network Architectures}
\label{appsec:arch}

\subsection{Discriminator}

\begin{table}[!htp]\centering
\caption{Discriminator network architecture. OC, KS, S, P, and NS stand for output channels, kernel size, stride, padding, and negative slope, respectively. The network architecture is adopted from~\cite{choi2017stargan,siddiquee2019learning} with slight modification.}
    \bgroup
    \def\arraystretch{1.5}
    \begin{tabular}{|P{1.8cm}|P{6cm}|P{2cm}|P{2cm}|}
        \hline
        \textbf{Type} & \textbf{Operations} & \textbf{Input Shape} & \textbf{Output Shape} \\
        \hline
        Input layer & Conv2d (OC=64, KS=4, S=2, P=1), LeakyReLU (NS=0.01) & $(h, w, 3)$ & $(\frac{h}{2}, \frac{w}{2}, 64)$ \\
        \hline
        \multirow{5}{1cm}{Hidden layers} & Conv2d (OC=128, KS=4, S=2, P=1), LeakyReLU (NS=0.01) & $(\frac{h}{2}, \frac{w}{2}, 64)$ & $(\frac{h}{4}, \frac{w}{4}, 128)$ \\
        & Conv2d (OC=256, KS=4, S=2, P=1), LeakyReLU (NS=0.01)  & $(\frac{h}{4}, \frac{w}{4}, 128)$ & $(\frac{h}{8}, \frac{w}{8}, 256)$ \\
        & Conv2d (OC=512, KS=4, S=2, P=1), LeakyReLU (NS=0.01)  & $(\frac{h}{8}, \frac{w}{8}, 256)$ & $(\frac{h}{16}, \frac{w}{16}, 512)$ \\
        & Conv2d (OC=1024, KS=4, S=2, P=1), LeakyReLU (NS=0.01) & $(\frac{h}{16}, \frac{w}{16}, 512)$ & $(\frac{h}{32}, \frac{w}{32}, 1024)$ \\
        & Conv2d (OC=2048, KS=4, S=2, P=1), LeakyReLU (NS=0.01) & $(\frac{h}{32}, \frac{w}{32}, 1024)$ & $(\frac{h}{64}, \frac{w}{64}, 2048)$ \\
        \hline
        Output layer ($D_{src}$) & Conv2d (OC=1, KS=3, S=1, P=1) & $(\frac{h}{64}, \frac{w}{64}, 2048)$ & $(\frac{h}{64}, \frac{w}{64}, 1)$ \\ \hline
    \end{tabular}
    \egroup
\end{table}

\newpage
\subsection{Generator}

\begin{table}[!htp]\centering
\caption{Generator network architecture. OC, KS, S, P, and IN stand for output channels, kernel size, stride, padding, and instance norm, respectively. The network architecture is adopted from~\cite{choi2017stargan,siddiquee2019learning} with slight modification.}
    \bgroup
    \def\arraystretch{1.5}
    \begin{tabular}{|P{1.5cm}|P{6cm}|P{2cm}|P{2cm}|}
        \hline
        \textbf{Type} & \textbf{Operations} & \textbf{Input Shape} & \textbf{Output Shape} \\
        \hline
        \multirow{3}{*}{Encoder} & Conv2d (OC=64, KS=7, S=1, P=3), IN, ReLU & $(h, w, 3)$ & $(h, w, 64)$ \\
        & Conv2d (OC=128, KS=4, S=2, P=1), IN, ReLU & $(h, w, 64)$ & $(\frac{h}{2}, \frac{w}{2}, 128)$ \\
        & Conv2d (OC=256, KS=4, S=2, P=1), IN, ReLU & $(\frac{h}{2}, \frac{w}{2}, 128)$ & $(\frac{h}{4}, \frac{w}{4}, 256)$ \\
        \hline
        \multirow{6}{*}{Bottleneck} & Residual Block: Conv2d (OC=256, KS=3, S=1, P=1), IN, ReLU & $(\frac{h}{4}, \frac{w}{4}, 256)$ & $(\frac{h}{4}, \frac{w}{4}, 256)$ \\
        & Residual Block: Conv2d (OC=256, KS=3, S=1, P=1), IN, ReLU & $(\frac{h}{4}, \frac{w}{4}, 256)$ & $(\frac{h}{4}, \frac{w}{4}, 256)$ \\
        & Residual Block: Conv2d (OC=256, KS=3, S=1, P=1), IN, ReLU & $(\frac{h}{4}, \frac{w}{4}, 256)$ & $(\frac{h}{4}, \frac{w}{4}, 256)$ \\
        & Residual Block: Conv2d (OC=256, KS=3, S=1, P=1), IN, ReLU & $(\frac{h}{4}, \frac{w}{4}, 256)$ & $(\frac{h}{4}, \frac{w}{4}, 256)$ \\
        & Residual Block: Conv2d (OC=256, KS=3, S=1, P=1), IN, ReLU & $(\frac{h}{4}, \frac{w}{4}, 256)$ & $(\frac{h}{4}, \frac{w}{4}, 256)$ \\
        & Residual Block: Conv2d (OC=256, KS=3, S=1, P=1), IN, ReLU & $(\frac{h}{4}, \frac{w}{4}, 256)$ & $(\frac{h}{4}, \frac{w}{4}, 256)$ \\
        \hline
        \multirow{3}{*}{Decoder} & ConvTranspose2d (OC=128, KS=4, S=2, P=1), IN, ReLU & $(\frac{h}{4}, \frac{w}{4}, 256)$ & $(\frac{h}{4}, \frac{w}{4}, 128)$ \\
        & ConvTranspose2d (OC=64, KS=4, S=2, P=1), IN, ReLU & $(\frac{h}{2}, \frac{w}{2}, 128)$ & $(h, w, 64)$ \\
        & ConvTranspose2d (OC=4, KS=7, S=1, P=3), Tanh & $(h, w, 64)$ & $(h, w, 4)$ \\
        \hline
    \end{tabular}
    \egroup
\end{table}

\newpage

\bibliographystyle{splncs04}
\bibliography{mybib}

\end{document}